\documentclass[pra,aps,epsfig,twocolumn,amsmath,amssymb,showpacs]{revtex4} 

\newcommand\be{\begin{eqnarray}}
\newcommand\ee{\end{eqnarray}}

\usepackage{graphicx} 

\begin{document}
\title{Decoherence-free generation of many-particle entanglement
by adiabatic ground-state transitions}
\author{R. G. Unanyan and M. Fleischhauer}
\affiliation{Fachbereich Physik der Universit\"{a}t Kaiserslautern, 67653\\
Kaiserslautern, Germany}
\date{\today }
 
\begin{abstract} We discuss a robust and decoherence insensitive method to
create many-particle entanglement in a spin system
with controllable collective interactions of the Lipkin-Meshkov-Glick type 
and propose an implementation in 
an ion trap. 
An adiabatic change of parameters allows a transfer from 
separable to a large variety of entangled eigenstates.  
We shown that under certain conditions the Hamiltonian
possesses a supersymmetry permitting an explicit construction of the
ground state at all times of the adiabatic transfer. 
Of particular interest is a transition in a {\it non-degenerate} ground state
with a finite energy gap since
here the influence of collective as well as individual
decoherence mechanisms is substantially reduced.
A lower bound for the energy gap is given.
\end{abstract} 
 
\maketitle 
 
Entanglement is one of the most characteristic features of quantum 
systems and lies at the heart of quantum  
information processing and computing \cite{entanglement}. While in  
few-particle systems entanglement is by now reasonably well understood  
\cite{entanglement-definition} and 
practical schemes for its creation and manipulation are developed 
\cite{few-particle-entanglement-experiments,WinelandScience2000},  
many-particle entanglement is still an open field of research with 
a large unexplored potential for applications.  
There exist for example proposals to implement quantum computation 
in an initially entangled many-particle system by performing only
measurements and 
single qubit operations, both of which are relatively easy to implement 
\cite{Briegel}. A necessary prerequisite for this is however a specific 
many-particle  
entanglement. Since entangled states become increasingly susceptible to 
environmental interactions if the number of particles increases, 
an important practical challenge is the design of 
robust and most importantly decoherence-resistant
mechanisms for its  generation. 
We here propose and analyse 
such a mechanism which is based on adiabatic transitions 
in a spin system with 
controllable collective interactions. 
Robustness against decoherence
and short process times are achieved by choosing ground-state 
transitions with a large energy gap very similar to the ideas
used in adiabatic quantum computation 
\cite{Preskill}.

Let us consider a collection of $N$ interacting spin $1/2$ systems
described by a generalization of 
the Lipkin-Meshkov-Glick (LMG) Hamiltonian
\cite{Lipkin1965}
%
%
\be
H=\xi\Bigl[
\lambda\chi_1\chi_2 \hat J_z +\chi_1^2 \hat J_x^2 +\chi_2^2 \hat J_y^2
+2\mu\chi_2^2 \hat J_y\Bigr],
\label{H-LMG}
\ee
where the $\hat J_i$, ($i\in\{x,y,z\}$) are the total-spin operators
of the ensemble. 
We show that this Hamiltonian can be implemented
for $\mu=0$ with controllable parameters
by a generalization of the ion-trap scheme suggested 
by S\o rensen and M\o lmer \cite{Sorensen1999}, where
cold ions interact via a common trap oscillation  
and are driven by bichromatic lasers fields. 
In contrast  to the
effective Hamiltonian of \cite{Sorensen1999} 
the generalized LMG interaction (\ref{H-LMG}) can 
not be solved exactly. It does provide however the
possibility for adiabatic and therefore robust transitions
between separable and entangled many-particle states.
The most important feature of (\ref{H-LMG}) is that such transitions 
are possible while staying in a {\it non-degenerate} 
ground state with a finite energy gap 
and that a large class of entangled target states are accessible by
varying $\mu$.
We show that in the special case of $\lambda=1$, the LMG Hamiltonian
(\ref{H-LMG}) has a 
supersymmetry and the ground-state can be explicitly constructed for all 
times of the transfer process.
The gap can be made rather large and hence the 
process can be fast despite of its adiabatic nature.
For the same reason the influence of decoherence is strongly
reduced:
{\it Collective} decoherence due to noise in the
external control parameters is eliminated by the adiabatic 
nature of the process and
the influence of {\it independent} individual reservoir 
couplings is suppressed due to the
presence of a finite energy gap.

Let us first discuss the case of a negative coupling parameter $\xi<0$.
If $\mu=0$ and $\lambda\ge N$,
the system described by eq.(\ref{H-LMG}) undergoes a quantum-phase 
transition \cite{Sahdev} when the
interaction parameters $\chi_1$ and $\chi_2$ are changed
(case {\it i}). 
This transition can be described analytically in the
semiclassical limit with $J=N/2$ and $N\gg 1$.
If $\chi_1=\chi_2$, $\hat J_z$ is a conserved
quantity and (\ref{H-LMG}) has a trivial anharmonic spectrum: 
%
%
$H\enspace \rightarrow \enspace 
\xi\chi_2^2\bigl[\lambda \hat J_z -\hat J_z^2 +\hat{\textbf{J}}^2\bigr].
$
%
%
The ground state of the system has maximum total angular momentum $J=N/2$
and a $z$ projection $m_z=\pm N/2$
depending on the sign of $\lambda$. 
Both of these states denoted by $|\uparrow\uparrow\dots\rangle$
and $|\downarrow\downarrow\dots\rangle$ are separable many-particle states.
On the other hand in the limit $\chi_1=0$ the terms
in (\ref{H-LMG}) containing ${\hat J}_x$ and ${\hat J}_z$
vanish and the Hamiltonian approaches 
%
%
$H\enspace \rightarrow \enspace \xi\chi_2^2
{\hat J}_y^2.\label{H_nonl}
$
%
%
It has again a trivial spectrum, however with two
{\it degenerate} ground states $|m_y=\pm N/2\rangle$.
Both of these, taken individually, are
separable. Symmetric or anti-symmetric superpositions of them form however
the $N$-spin analogues of the Greenberger-Horne-Zeilinger states (GHZ
states). 
Using the Schwinger-representation of angular momenta
\cite{Schwinger}  one finds that the state 
$|m_z=N/2\rangle$ is adiabatically connected only to one particular
superposition due to the symmetry of the Lipkin interaction. Thus 
\begin{equation}
|\uparrow\uparrow\dots\rangle \quad\stackrel{(i)}{\longrightarrow}\quad 
\frac{1}{\sqrt{2}}\Bigl[|++\dots\rangle + {\rm e}^{i\pi J}
|--\dots\rangle\Bigr],  \label{transform_i}
\end{equation}
which corresponds to the generation of the $N$-particle analog of the GHZ
state in the $\sigma_y$ basis.
$|\pm\rangle$ denote single-particle eigenstates of $\sigma_y$ 
with $m_y=\pm 1/2$ respectively.
Due to the phase factor $e^{i\pi J}$ the entangled state depends 
sensitively on the total
number of atoms even in the limit $N\rightarrow \infty $ \cite{Pokrovsky}.

For a positive coupling parameter, $\xi>0$, three cases need
to be distinguished: $\mu=0$ and the total
number of spins is odd (case {\it ii}); $\mu=0$ and the total
number of spins is even (case {\it iii}); and $\mu\ne 0$ 
(case {\it iv}). In all cases the ratio $\chi_1/\chi_2$ is 
again rotated from
unity to zero.

Since $\xi>0$ the initial ground state in cases ({\it ii}) and ({\it iii})
is $|m_z=\pm N/2\rangle$ depending on the sign of $\lambda\ne 0$ 
and is separable. 
The final Hamiltonian is again $H=\xi\chi_2^2\hat J_y^2$, whose
ground state is now however the eigenstate of ${\hat J}_y$ 
with smallest value of $|m_y|$. Thus for an odd number 
of particles (case {\it ii}) there are two
{\it degenerate} ground states with $m_y=\pm 1/2$, both of them being
maximally entangled. Making use of the symmetry 
of the interaction and using the Schwinger-representation 
we find that the 
adiabatic transition leads to the mapping 
\begin{equation}
|\downarrow \downarrow \dots \rangle \,\stackrel{(ii)}{\longrightarrow 
}\,\frac{1}{\sqrt{2}}\Bigl[\bigl|(+)^{n}(-)^{n-1}\bigr\rangle_{{\rm s}%
}+i\bigl|(+)^{n-1}(-)^{n}\bigr\rangle_{{\rm s}}\Bigr],
\end{equation}
where $n=\lceil N/2\rceil $ and the subscript ``s'' denotes symmetrization.

In case ({\it iii}) the final ground
state ($\chi_1\to 0$) is {\it non-degenerate}, has
spin projection $m_y=0$ and is maximally entangled.
\begin{equation}
|\downarrow\downarrow\dots\rangle \quad\stackrel{(iii)}{\longrightarrow}%
\quad |m_y=0\rangle \, =\, \bigl|(+)^{N/2}(-)^{N/2}\bigr\rangle _{{\rm s}}.
\end{equation}
In this case there is no merging of eigenstates and consequently 
no phase transition. The absence of degeneracy during the
entire adiabatic transfer makes case ({\it iii}) particularly interesting
because here decoherence is strongly suppressed by the presence of a finite
energy gap. The target state is however fixed to the special case
$|m_y=0\rangle$.

The variety of accessible target states in a non-degenerate adiabatic 
ground-state transition can be substantially increased by adding a
linear interaction proportional to $\hat J_y$ to the LMG
Hamiltonian, i.e. by allowing for a non-vanishing value of $\mu$ in 
eq.(\ref{H-LMG}), (case {\it iv}): If we assume for simplicity
that $\mu=m$, with $m$ being an integer or half-integer
with $m\in\{-J,-J+1,\dots,J\}$, the final ($\chi_1\to 0$) 
Hamiltonian approaches $
H\enspace \rightarrow \enspace \xi\chi_2^2\Bigl[
{\hat J}_y^2+2 m {\hat J}_y\Bigr]$.
Its  ground state  is $|m_y=m\rangle$ and thus 
can be adjusted to 
have {\it any} eigenvalue of ${\hat J}_y$,
which is maximally entangled unless $m=\pm J$.
\begin{equation}
|\downarrow\downarrow\dots\rangle \quad\stackrel{(iv)}{\longrightarrow}%
\quad |m_y=m\rangle.
\end{equation}

\begin{figure}[ht]
\includegraphics[width=8cm]{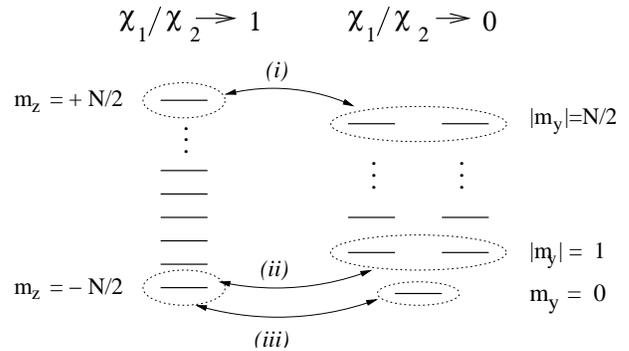} 
\caption{Energy spectrum of Hamiltonian (\ref{H-LMG}) for $\xi>0$, even
number of particles $N$ and maximum $J$ in the limits for $\chi_1/\chi_2\to 1$
and $\lambda>0$ (left) and $\chi_1/\chi_2\to 0 $ (right). 
If the sign of $\xi$ is
changed the picture has to be flipped upside down and if the number of spins
is odd (case ({\it ii})) the state $m_y=0$ does not exist 
and the state $|m_y|=1$ needs
to be replaced by $|m_y|=1/2$. 
Adiabatic change of $\chi_1/\chi_2$ 
allows for robust transfer between separable (left) and entangled
eigenstates (right) following the possible scenarios ({\it i-iii}).}
\label{fig2}
\end{figure}

In all four cases considered an adiabatic
change of $\chi_1/\chi_2$ from unity to zero
leads to the generation of a well defined entangled state. 
 In cases
({\it iii}) and ({\it iv}) this is moreover possible with a finite
energy gap. The adiabatic
process is neither of the Landau-Zener nor the STIRAP type
\cite{VitanovAdvAMO2001}.
We have illustrated the different scenarios for even $N$, $\mu=0$,
 and a positive linear
term in (\ref{H-LMG}) in Fig. \ref{fig2}.

To find the ground-state of (\ref{H-LMG}) for arbitrary values of 
$\chi_1/\chi_2$ as e.g. during the transfer process is a difficult task.
We will show now that (\ref{H-LMG}) possesses a 
super-symmetry (SUSY) for $\lambda=1$,
allowing for an explicit construction of the ground state for 
cases ({\it iii}) and ({\it iv}). (The existence of an extra symmetry 
has been suspected \cite{GargPRB2001} but has not been understood so far.)
For $\lambda=1$ the Hamiltonian can be factorized as
\be
\frac{H}{\xi}&&=
\left(\chi_1\hat J_x+i\chi_2\hat J_y-i\chi_2\mu\right)
\left(\chi_1\hat J_x-i\chi_2\hat J_y+i\chi_2\mu\right)\nonumber\\
&& -\chi_2^2\mu^2.
\ee
Due to the SUSY the spectrum
consists of twofold degenerate and
a nondegenerate state \cite{WittenNuclPhys1981}. 
For $\mu=m\in\{-J,\dots,J\}$ the ground state is nondegenerate and
obeys $(\chi_1\hat J_x-i\chi_2\hat J_y+i\chi_2 m)\, |\psi_0\rangle =0$.
One finds
\be
|\psi_0\rangle ={\cal N}\, \exp\left(\gamma \hat J_z\right)\, 
|m_y=m\rangle\label{ground-state-gen}
\ee
where ${\cal N}$ is a normalization constant and $\tanh(\gamma) =\chi_1/\chi_2$
for $\chi_1\le \chi_2$. $|\psi_0\rangle$ is an entangled state of the $N$
spins for any value of $\gamma\ne 0$. I.e. by varying $\gamma$ and 
choosing $m$, we have access
to a rich variety of entangled many-particle states while staying in
the lowest energy state.

We proceed by discussing  the conditions for adiabaticity
and the sensitivity of the process to decoherence.
In this context particular attention
has to be given to the phase transitions in cases ({\it i}) and ({\it ii}) 
since they are associated with a merging of pairs of energies.
Due to the symmetry
of (\ref{H-LMG}) only one superposition of the associated
states is however coupled
to the nondegenerate initial ground state.
Consequently for the transition to be adiabatic it is sufficient that the
characteristic time of the transfer $T$ is much larger than the typical
inverse frequency difference $\hbar/\Delta E$ 
to the next excited states. For the same reason
{\it collective} decoherence processes caused e.g. by fluctuations in the
external parameters $\xi,\lambda,\chi_1$, and $\chi_2$ are suppressed
in the present system by an exponential Boltzmann factor 
$\exp(-\beta \Delta E)$, where $\beta^{-1} =k_B T$ is the thermal energy
of the heat bath. 
Decoherence processes caused by {\it independent} heat bath couplings of the 
individual spins do not have the symmetry of (\ref{H-LMG})
 however and hence do couple the degenerate
states in cases ({\it i}) and ({\it ii}). 
Only in cases ({\it iii}) and ({\it iv}), 
i.e. without a quantum phase transition, there is
always a finite energy gap to all other states with the same total angular
momentum $J$. 
Thus the 
entanglement generation
in cases ({\it iii}) and ({\it iv}) will be robust against collective {\it and}
individual decoherence processes, provided the energy gap is sufficiently 
large.

It is not possible to give an analytic expression for the energy gap 
$\Delta E$ in the
general case. Numerical investigations for up to 50 particles indicate
that for a transfer efficiency close to 100\% it is sufficient that 
\begin{equation}
\chi _{1}^{{\rm max}}\,\sqrt{T\,}\sim \,\chi _{2}^{{\rm max}}\,\sqrt{T\,}
\gg 1
\end{equation}
with $\lambda \ge N$ for case ($i$) and $\lambda \ge 1$ for cases 
($ii-iv$). $T$ is the characteristic transfer time.
An estimate for $\Delta E $ in cases ({\it ii}) and ({\it iii}),
i.e. for $\mu=0$ can be obtained as follows: Using a
variational method with a trial functions $|\Psi ^{(N)}\rangle $ one finds
an energy estimate $\langle H\rangle _{N}$ for the ground state with $J=N/2$, 
$E_{0}^{(N)}\leq \langle H\rangle _{N}$. Secondly one can apply
Temple's formula \cite{Temple52} to obtain a lower bound 
$E_{0}^{(N)}\geq \langle
H\rangle _{N}-\frac{\langle \Delta H^{2}\rangle _{N}}{E
_{1}^{(N)}-\langle H\rangle _{N}}$, where $\langle \Delta H^{2}\rangle
_{N}=\langle \bigl(H-\langle H\rangle \bigr)^{2}\rangle _{N}$ is the energy
fluctuation in the trial state, $E_{1}^{(N)}$ is the energy of the 
first excited state. One can show that Temple's formula gives the
best lower bound when using only  $\langle \Delta H^{2}\rangle _{N}$, 
$\langle H\rangle _{N}$  and $E_{1}^{(N)}$ as parameters. 
Furthermore we make use of the inequality $E_{1}^{(N)}\geq E
_{0}^{(N-2)}$ between the energy of the first excited state for $N$
particles and the ground state for $N-2$ particles, in both cases with
maximum $J$, which can easily be proven. Applying this inequality
iteratively leads to  
\begin{equation}
\frac{\Delta E^{(N)}}{\langle H\rangle _{N-2}-\langle H\rangle _{N}}\geq 
\frac{1}{2}\left( 1+\sqrt{1-4\,A}\right) 
\end{equation}
where 
\begin{equation}
A>\frac{\langle \Delta H^{2}\rangle _{N}}{\left( \langle H\rangle
_{N-2}-\langle H\rangle _{N}\right) \left( \langle H\rangle _{N-4}-\langle
H\rangle _{N-2}\right) }.
\end{equation}
By choosing the trial function of the ground state close to the exact one it
is possible to achieve $4\,A\ll 1$, i.e. $\Delta E^{(N)}\sim \langle H\rangle
_{N-2}-\langle H\rangle _{N}$. 
 For the
simple trial function $|\Psi ^{(N)}\rangle =\alpha _{1}|m_{y}=0\rangle
+\alpha _{2}|m_{z}=-N/2\rangle $ one finds $\langle H\rangle _{N-2}-\langle
H\rangle _{N}=\beta \lambda \left| \xi \right| \chi _{1}\chi _{2}
\vert_{\rm max}$, where $%
\beta $ is a numerical factor of order unity, which varies only very slowly
with $N$. Thus a reasonable estimate for the energy gap in case ($iii$) is
given by $\beta \lambda \left| \xi \right| \chi _{2}^2$. This is also
confirmed by our numerical calculations for particle numbers up to $N=50$.
If $\beta \lambda \left| \xi \right| \chi _{2}^2$ is sufficiently
larger than the thermal energy of the environment, the probability of
decoherence processes is strongly suppressed.

\begin{figure}[ht]
\includegraphics[width=7cm]{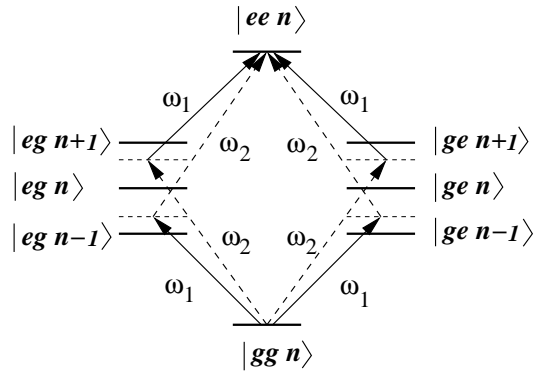} 
\caption{Excitation scheme of pair of ions with ground state $|g\rangle$ and
excited state $|e\rangle$ by bichromatic laser fields of equal and opposite
detuning $\protect\delta$. $n$ denotes quantum number of trap oscillation.}
\label{fig1}
\end{figure}

Let us now discuss a possible implementation of the Lipkin Hamiltonian
(\ref{H-LMG}) with $\mu=0$ in an ion-trap system. 
Consider a linear trap with a string of ions with two relevant
internal levels $|g\rangle$ and $|e\rangle$. The ions are assumed to be
cooled such that only the  
in-phase collective oscillation of all ions is excited. 
The corresponding oscillation
frequency is denoted by $\nu$. The two internal levels of the ions are
coupled by two laser fields with frequencies $\omega_1$ and $\omega_2$, and
slowly-varying Rabi-frequencies $\Omega_1$ and $\Omega_2$. Assuming that
both fields couple all ions in the same way, we can describe the system by
the Hamiltonian 
$H=H_{0}+H_{{\rm int}}$, where 
$H_{0}=\hbar \nu \hat c^{\dagger }\hat c +\hbar \omega _{eg}\hat J_{z}$,  
$\hat c$ and $\hat c^\dagger$ being the annihilation and creation operators
of the trap oscillation. $\hbar\omega_{eg}$ is the energy separation between
the two internal states and $\hat J_z=\frac{1}{2} \sum_{i=1}^N
(\sigma_{ee}^i-\sigma_{gg}^i)$, with $\sigma^i_{\mu\mu}=
|\mu\rangle_{ii}\langle\mu|$ being the projector to the internal state $%
|\mu\rangle$ of the $i$th ion. The interaction Hamiltonian $H_{{\rm int}}$
is given in rotating wave approximation by
\begin{equation}
H_{{\rm int}}=\hat J_{+} {\rm e}^{i\eta (\hat c+\hat c^{\dagger })} \Bigl[%
\Omega _{1}{\rm e}^{ -i\omega_{1}t } +\Omega _{2} {\rm e}^{-i\omega _{2}t}
\Bigr]+h.c.  \label{H_int}
\end{equation}
where $\eta$ is the Lamb-Dicke parameter. We assume that the laser
frequencies have equal and opposite detuning $\delta$ from resonance $\omega
_{1,2}=\omega _{eg}\pm \delta$. $\delta$ is assumed to be large
compared to the linewidth of the resonance but sufficiently different from
the frequency of the trap oscillation, i.e. $|\delta|,|\nu\pm\delta|\gg
\gamma$. As a consequence the dominant processes are two-photon transitions
leading to a simultaneous excitation of pairs of ions as indicated in Figure 
\ref{fig1}. For $\Omega_1=\Omega_2$ this scheme has first been considered by
S\o rensen and M\o lmer \cite{Sorensen1999,Molmer1999} in the context of
quantum computation and dynamical entanglement generation.

\begin{figure}[ht]
\includegraphics[width=8cm]{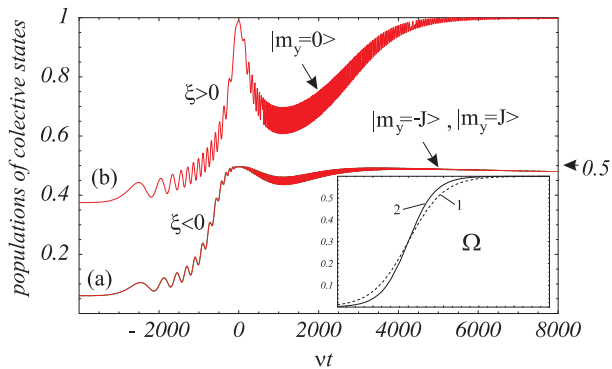} 
\caption{Numerical simulation of the adiabatic transfer for
a system of four ions. Shown are 
populations in states $|m_y=0\rangle$ and $|m_y=\pm J\rangle$ for
$\Omega _{1,2}\left( t\right) =(\alpha/2)\left( \tanh(t/T_{1,2})%
+1\right)$ as shown in the inset and for 
$\alpha /\nu =0.6,\nu T_{1}=2000,\nu T_{2}=1500,$ and $%
\delta /\nu =0.9$\,  (a), case $\left( i\right) $; and $\delta /\nu =1.1$
\, (b), case $\left( iii\right) .$}
\label{fig3}
\end{figure}

We now assume that the ion trap is in the Lamb-Dicke limit, i.e. that the
ions are cooled sufficiently enough, such that for all relevant excitation
numbers $n$ of the trap oscillation $(n+1)\eta^2 \ll 1$ holds. In this limit
one can expand the exponent in (\ref{H_int}) to first order in $\eta$. 
For large values of $|\delta| $ it is convenient to consider this
interaction in terms of a coarse-grained Hamiltonian which neglects the
effects of rapidly oscillating terms. 
Using the time-averaging method of ref.
\cite{James} one arrives at an effective  Lipkin Hamiltonian (\ref{H-LMG})
with the identifications $\xi = (2\nu\eta^2)/({\delta^2-\nu^2})$,
$\lambda = {2}/{\xi\delta}$, and
$\chi_{1,2} = \Omega_1\mp\Omega_2$.
$\Omega_1=\Omega_2$ corresponds 
to the case $\chi_1=\mu=0$ in (\ref{H-LMG}) which is exactly
solvable and has been discussed in Refs.\cite{Sorensen1999,Molmer1999}.
It has also been shown in \cite{Sorensen1999,Molmer1999} that the 
effective Hamiltonian describes correctly the dynamics of the ions in 
a coarse-grained time picture. Furthermore the coupling scheme has
successfully been implemented in experiments to generate entanglement
between 4 ions \cite{WinelandScience2000}.

In Fig.\ref{fig3} we have shown as an example the effective 
dynamics of a system of 4 ions driven by fields $\Omega_1$ and $\Omega_2$
for 
the cases ({\it i}) and ({\it iii}) following eq.(\ref{H-LMG}) with $\mu=0$.
 One recognizes that a nearly perfect
transfer is possible.

In summary we have shown that it is possible to generate specific
entangled many-particle states in an ensemble of spins interacting 
through a collective coupling of the Lipkin-Meshkov-Glick type
by adiabatic ground-state transitions. 
Two scenarios ({\it i,ii}) involve a two-fold degenerate ground state
at some stages 
while two others ({\it iii}) and ({\it iv}) 
always have a nondegenerate ground state. In all cases 
there is a finite energy gap to other excited states.
This gap can be rather large and thus fast processes 
are possible despite the required adiabaticity.
In the asymptotic limits of the adiabatic transfer, the
spectrum and eigenstates of the LMG-Hamiltonian can be exactly
calculated. Furthermore for $\lambda=1$ 
there is a supersymmetry allowing for an 
explicit construction of the ground state for all times.
Due to adiabaticity, all transitions are robust against parameter
variations. Furthermore due to the symmetry of the coupling and the
finite energy gap from the (degenerate or nondegenerate) ground state
to excited states, {\it collective} decoherence processes are suppressed. 
In addition in the case of a nondegenerate ground state (case {\it iii} and
{\it iv})
also {\it individual} decoherence processes are suppressed. 
The collective spin Hamiltonian can be implemented
in a cold ensemble of ions in a linear trap
driven by a nearly-resonant bichromatic fields.


The authors would like to thank C. Mewes and L. Plimak for their help 
with numerics.  This work
has been supported by the Deutsche Forschungsgemeinschaft 
under contract no. Fl-210/10. 
RU also acknowledges financial support 
by the Alexander-von-Humboldt foundation.



\begin{references}
\bibitem{entanglement}  D. Bouwmeester, A. Ekert, and A. Zeilinger, {\it The
Physics of Quantum Information: Quantum Cryptography, Quantum
Teleportation, Quantum Computation} (Springer-Verlag, Berlin, 2000).

\bibitem{entanglement-definition} W.K. Wooters, Phys. Rev. Lett. {\bf 80}, 
2245 (1998).

\bibitem{few-particle-entanglement-experiments}  C. Monroe {\it et al}.,
Phys. Rev. Lett. {\bf 75}, 4714 (1995); E. Hagley {\it et al}., Phys. Rev.
Lett. {\bf 79}, 1 (1997).

\bibitem{WinelandScience2000} C.A. Sackett, {\it et al.}
Nature (London), {\bf 404}, 256 (2000).

\bibitem{Briegel}  R. Raussendorf and H.J. Briegel, Phys. Rev. Lett. {\bf 86%
}, . 5188 (2001).

\bibitem{Preskill}  A. M. Childs, E. Farhi and J. Preskill, Phys. Rev. A 
{\bf 65}, 012322 (2002). 


\bibitem{Lipkin1965}  H.J. Lipkin, N. Meshkov and A. Glick, Nucl. Phys. {\bf %
62}, 188  (1965).

\bibitem{Sorensen1999}  A. S\o rensen and K. M\o lmer, Phys. Rev. Lett. {\bf %
82}, 1971 (1999).


\bibitem{Sahdev}  S. Sachdev {\it ''Quantum Phase Transitions''},
(Cambridge-Univ. Press, Cambridge, 1999).

\bibitem{Schwinger} L.C. Biedenharn, J.D. Lauck, {\it `` Angular momentum in
quantum physics: Theory and applications''} (Addison-Wesley, Reading 1981).




\bibitem{Pokrovsky}  V.A. Kalatsky, E. M\"{u}ller-Hartmann, V.L. Pokrovsky,
and G.S. Uhrig, Phys. Rev. Lett. {\bf 80}, 1304 (1998).


\bibitem{VitanovAdvAMO2001} N.V. Vitanov, {\it et al.}, 
Adv. Atom. Mol. Phys.  Vol.{\bf 46}, 55-190 (Academic Press, 2001).

\bibitem{GargPRB2001} A. Garg, Phys. Rev. B {\bf 64}, 094413 (2001).

\bibitem{WittenNuclPhys1981} E. Witten, Nucl. Phys. B, {\bf 188}, 513 (1981).

\bibitem{Temple52}  G. Temple, Proc. Royal Soc. A, {\bf 211}, 204 (1952).

\bibitem{Molmer1999}  K. M\o lmer and A. S\o rensen,  Phys. Rev. Lett. {\bf %
82}, 1835 (1999). 

\bibitem{James}  D. F. James, Fortschr. Phys. {\bf 48}, 9 (2000).



\end{references}
\end{document}